\documentclass[aps,prd,preprintnumbers]{revtex4}

\usepackage{epsfig}
\usepackage{bm}
\newcommand{\vect}[1]{\bm{#1}}  
\newcommand{\text}{\mathrm}
\newcommand{\mq}{|\vect{q}|}
\newcommand{\MeV}{\text{MeV}}
\newcommand{\keV}{\text{keV}}
\newcommand{\GeV}{\text{GeV}}
\newcommand{\grad}{\mathbf{\nabla}}
\newcommand{\tmu}{\tilde{\mu}}
\newcommand{\CG}[6]{\left<{#3},{#6}|{#1},{#4};{#2},{#5}\right>}  

\newcommand{\be}{\begin{equation}}
\newcommand{\ee}{\end{equation}}
\newcommand{\bea}{\begin{eqnarray}}
\newcommand{\eea}{\end{eqnarray}}
\newcommand{\bean}{\begin{eqnarray}}
\newcommand{\eean}{\end{eqnarray*}}

\newcommand{\gapproxeq}{\lower
.7ex\hbox{$\;\stackrel{\textstyle >}{\sim}\;$}}
\newcommand{\lapproxeq}{\lower
.7ex\hbox{$\;\stackrel{\textstyle <}{\sim}\;$}}

\def\3bar{$\bar {\hbox{\bf 3}}$}

\begin{document}

\title{Is X(3872) a molecule?}
\author{C. E. Thomas}
\email[E-mail: ]{c.thomas1@physics.ox.ac.uk}
\author{F. E. Close}
\email[E-mail: ]{f.close1@physics.ox.ac.uk}
\affiliation{Rudolf Peierls Centre for Theoretical Physics, University of Oxford,\\ 1 Keble Road, Oxford, OX1 3NP}
\date{23 May 2008}
\preprint{OUTP-08-08P}

\begin{abstract}
We show that the literature on pion exchange between charm and bottom mesons is inconsistent.  We derive the formalism explicitly, expose differences between papers in the literature and clarify the implications.  We show that the $X(3872)$ can be a bound state but that results are very sensitive to a poorly constrained parameter.  We confirm that bound states in the $B\bar{B}$ sector are possible. The circumstances whereby exotic combinations can bind with $cc$ or $bb$ quantum numbers are explored.
\end{abstract}

\maketitle

\section{Introduction}
\label{sec:Introduction}

The nature of the enigmatic charmonium meson $X(3872)$, which appears at $D^0D^{*0}$ threshold with $J^P = 1^+$, has been the subject of intense debate ever since its discovery.  Historically Ericson and Karl\cite{Ericson:1993wy} considered pion exchange in hadronic molecules.  Tornqvist predicted\cite{Tornqvist:1991ks,Tornqvist:1993ng} that one pion exchange between charmed mesons gives an attractive force in the $I=0$ channel such that molecular or resonant $D\bar{D}^*$ states might arise near threshold. Following the discovery of the $X(3872)$ several papers suggested that it could be a $1^{++}$ state, driven by pion exchange\cite{Close:2003sg,Tornqvist:2004qy} and/or quark exchange\cite{Swanson:2003tb} where flavour symmetry breaking was associated with its affinity for the neutral $D^0 D^{*0}$ threshold.

More recently several papers have appeared assessing the potential attractive forces and asking whether a bound state is dynamically realisable. These notably include Suzuki\cite{Suzuki:2005ha} who has argued that the one pion exchange forces are only able to make a feeble attraction at best, and most recently Liu et.\ al.\cite{Liu:2008fh,Liu:2008du} who have claimed that a bound state does not exist for reasonable values of parameters.

All of these papers\cite{Tornqvist:1993ng,Close:2003sg,Swanson:2003tb,Suzuki:2005ha,Liu:2008fh,Liu:2008du} make different assumptions of detail, are not always self-consistent, and do not all agree on the mathematical expressions even where their assumptions are the same. Hence the purpose of the present paper is to attempt a unified treatment of this problem, enabling comparison between the various approaches to be made. In particular we shall make explicit the calculation of some critical signs, upon which attraction or repulsion can depend, and whose derivation is not described in the existing literature. As a result we shall find that expressions in the various papers are mutually incompatible. We shall then propose a consistent formulation, discuss its consequences and compare with the existing literature.

Tornqvist\cite{Tornqvist:1993ng} initially assumed isospin symmetry and found $I=0$ attraction. Following the experimental discovery, Close and Page\cite{Close:2003sg} showed that the $d-u$ mass difference can lead to substantial breaking of the flavour symmetry enabling attraction in the neutral $D^0 \bar{D}^{*0}$ configuration.  Tornqvist also studied isospin symmetry breaking\cite{Tornqvist:2004qy}.  Recently Liu et.\ al.\cite{Liu:2008fh,Liu:2008du}, using the empirical fact that the state is at neutral threshold, have focused solely on the neutral channel without discussion of isospin symmetry or its breaking.  In addition, they assume that any phenomenon at the $BB^*$ threshold also involves only one charge channel. As we shall discuss here, the mechanism of flavour symmetry breaking can be critical in deciding which channels if any are attractive, and in particular whether attractive forces are strong enough to bind.  We will also note that the $B\bar{B}^*$ and $D\bar{D}^*$ situations can be very different. As we shall argue, the symmetry breaking and dynamics in the $D\bar{D}^*$ relative to $B\bar{B}^*$ cases depend on the mass splittings between vector and pseudoscalar masses and whether they are larger or smaller than the $\pi$ mass.  This has been noted clearly in the work of Suzuki\cite{Suzuki:2005ha} and Liu et.\ al.\cite{Liu:2008fh,Liu:2008du} where the Fourier transform gives different potentials in position-space, but is not apparent in the original work of Tornqvist\cite{Tornqvist:1993ng}. Furthermore, we find differences in some critical signs relative to Tornqvist in Ref.\ \cite{Tornqvist:1993ng}.  These have potential implications for the attraction or repulsion in the $D\bar{D}^*$ and $B\bar{B}^*$ systems which differ from that reference.

In the present paper we shall first derive the expression for the $\pi$-exchange potential along the lines of the original paper\cite{Tornqvist:1993ng}.  This will expose the origin of the signs that determine the overall attraction and repulsion and how the vector-pseudoscalar mass gap is critical. We shall concentrate on making contact with existing literature, showing where there are differences of assumption, sensitivity to inputs, and possible errors of calculation.  Finally we shall assess the implications.

In Section \ref{sec:ChargeConj} we make pedagogic comments about different conventions for charge conjugation eigenstates, in order to clarify discussions in the literature and to define our formalism.  In Section \ref{sec:QuarkStates} we give a simple  illustration of the spin expectation values for $P\bar{V}$ ($V\bar{P}$) and $V\bar{V}$.  This exposes a relative sign between these  that disagrees with Ref.\ \cite{Tornqvist:1993ng}.  In Section \ref{sec:OverallSign} we give the overall spin and flavour factor.   We calculate the effective potential in position space in Section \ref{sec:PositionSpacePotential}; our result in equation (23) exposes the differences in the existing literature\cite{Tornqvist:1993ng,Suzuki:2005ha,Liu:2008fh,Liu:2008du}.

In Section \ref{sec:Applications} we move on to applications of the formalism: we discuss normalisation in Section \ref{sec:Normalisation}, show the shape of the potentials in Section \ref{sec:PlotPotentials}, and apply the formalism to the $D\bar{D}^*$ system in Section \ref{sec:DDv}.  In Section \ref{sec:FurtherApps} we study the $B\bar{B}^*$  and flavour exotic $DD^*$ and $BB^*$ systems.   We finish with some general comments and conclusions in Section \ref{sec:Conclusions}.

\section{Basics}
\label{sec:Basics}

The effective one pion exchange potential between two light quarks can be split into a central term proportional to $(\vect{\sigma}_i \cdot \vect{\sigma}_j)(\vect{\tau}_i \cdot \vect{\tau}_j)$ and a tensor term proportional to $S_{ij}(\vect{\hat{r}})(\vect{\tau}_i \cdot \vect{\tau}_j)$ with $S_{ij}(\vect{\hat{q}}) \equiv 3(\vect{\sigma}_i \cdot \vect{\hat{q}})(\vect{\sigma}_j \cdot \vect{\hat{q}}) - (\vect{\sigma}_i \cdot \vect{\sigma}_j)$.  $\vect{\tau}_i$ are the isospin matrices acting on light quark $i$ and $\vect{\sigma}_i$ are the Pauli spin matrices acting  on light quark $i$.  The matrix elements of these operators capture all the spin and flavour dependence.  The remaining  dependence on kinematics, normalisation of the potential and more detailed model assumptions is discussed in Sections \ref{sec:PositionSpacePotential} and \ref{sec:Applications}.

We obtain the effective potential between two hadrons by summing over all the interactions between light $u$ and $d$ quarks/antiquarks.  In heavy-light mesons such as $D$ and $B$ there is only one possible interaction to consider.

\subsection{Charge conjugation and conventions}
\label{sec:ChargeConj}

There has been some confusion in the literature as to the correct definition of the $C=+$ state made of $D\bar{D^*} \pm c.c.$.  In Ref.\ \cite{Tornqvist:1993ng} the state with definite $C$ parity is defined by $(P\bar{V})_{\pm} = [P\bar{V} \pm C(P\bar{V})]$.  As $C^2 \equiv 1$ the above equation is self-consistent but does not specify the wavefunction until one chooses a convention whether $CV(P) = \pm \bar{V}(\bar{P})$. Only the neutral state is an eigenstate of $C$ and so one has $CV = \pm \bar{V} = -V$, which is consistent with  $C\bar{V} = \pm V = -\bar{V}$.  The intermediate step $CV(P) = \pm \bar{V}(\bar{P})$ is arbitrary (i.e. convention dependent).

In Ref.\ \cite{Tornqvist:1991ks,Swanson:2003tb} the $C=+$ state is defined to be $D\bar{D^*} + \bar{D}D^*$, whereas in Refs.\ \cite{Liu:2008fh,Liu:2008du,Dong:2008gb} it is claimed that the `correctly argued' form is $D\bar{D^*} - \bar{D}D^*$. As the overall sign of attraction versus repulsion depends on this sign, it is important to understand the origin of these alternate forms. As a particular example of our discussion above, the $D$ and $D^*$ are not eigenstates of $C$, and so the eigenvalues for the eigenstates $D\bar{D^*} \pm \bar{D}D^*$ depend on what conventions are used to define the states.

Hence first we define our flavour states.  For a $q$ and $\bar{q}$ of given flavour and spin at positions 1 and 2, eigenstates of $C$ for $P \equiv 0^{-+}$ and $V \equiv 1^{--}$ are
\bea
P: (q(1)\bar{q}(2) + \bar{q}(1)q(2) )/\sqrt{2} \nonumber \\
V: (q(1)\bar{q}(2) - \bar{q}(1)q(2) )/\sqrt{2}
\eea
For a meson made of a heavy quark, $Q$, and light flavour $q$, we define the meson and `anti-meson' to be
\bea
~P: ( q(1)\bar{Q}(2) + \bar{Q}(1)q(2) )/\sqrt{2};~~~ \bar{P}: ( \bar{q}(1)Q(2) + Q(1)\bar{q}(2) )/\sqrt{2} \nonumber \\
V: ( q(1)\bar{Q}(2) - \bar{Q}(1)q(2) )/\sqrt{2};~~~ \bar{V}: ( \bar{q}(1)Q(2) - Q(1)\bar{q}(2) )/\sqrt{2}
\eea
With these definitions the $C$-eigenstates are
\be
C|P\bar{V} \pm \bar{P} V  \rangle = \pm |P\bar{V} \pm  \bar{P} V \rangle. 
\ee

As Tornqvist \cite{Tornqvist:1991ks} used this convention, we shall do so in order to make most immediate comparison
with his results. We agree however that there are possible advantages in using the opposite convention, for example, in making contact 
with approaches that use interpolating currents in QFT
as argued in Refs.\ \cite{Liu:2008fh,Liu:2008du,Dong:2008gb}.

\subsection{Quark states and one pion exchange}
\label{sec:QuarkStates}
 
Our plan in this section is to calculate the sign of the spin expectation values of the central term in two cases: (i) $P\bar{V}$ or $V\bar{P}$ in $J_z=+1$ and (ii) $V\bar{V}$ in $J_z = 2$.  We will show that the $C=+$ combination of $(P\bar{V} + V\bar{P})$ has the {\it same} sign for the spin operator as does $V\bar{V}$ in $J_z = 2$.  This is opposite to the results in Tables 2 and 5 of Ref.\ \cite{Tornqvist:1993ng}.

As the $\pi$ vertex connects $D \leftrightarrow D^*$ (antiparticles understood here also) we
will calculate

\bea
\label{equ:PVHVP}
\langle P\bar{V} \pm \bar{P} V  | H |P\bar{V} \pm \bar{P} V  \rangle = 
\eta \langle P\bar{V} \pm \bar{P} V  | H |\bar{V} P \pm V \bar{P}  \rangle 
\eea
\noindent so that the sign of the effective matrix element will be given by

\be
\label{equ:PVVP}
\pm \eta \langle P| h | V  \rangle \langle \bar{V} | h | \bar{P}  \rangle 
\ee
The overall sign is therefore dependent on $\eta$.  

From general arguments, we have $\eta = 1$ for the $J^{P}=1^{+}$ state: because $L$ is even there is no phase change from the spatial wavefunction on swapping $\bar{V}(\bar{P})$ and $P(V)$.  There is no phase from the spin wavefunction because spin 1 coupling with spin 0 to give spin 1 has the same sign as spin 0 coupling with spin 1 to give spin 1.  To show this explicitly we need to define the quark
content of the mesons, including their spin orientation. For simplicity we consider the vector to be in
state $J_z = +1$, and for shorthand write
$[\bar{Q} q] \equiv \bar{Q} ^\uparrow q ^\uparrow$ (and similar for $\bar{q} Q$ etc), while the $S=0$
state is denoted 
$(\bar{Q} q) \equiv \sqrt{\frac{1}{2}}(\bar{Q} ^\uparrow q ^\downarrow - \bar{Q} ^\downarrow q^\uparrow)$ .

This implies that $|P\bar{V} \rangle$ and $| V \bar{P} \rangle$ are as follows:

\bea
|P\bar{V} \rangle = (\bar{Q} q)[\bar{q} Q] - (\bar{Q} q)[Q \bar{q} ] - (q \bar{Q} )[Q \bar{q} ] + (q \bar{Q} )[\bar{q} Q]
\eea

and

\bea
|V \bar{P} \rangle = - [\bar{Q} q](\bar{q} Q) - [\bar{Q} q](Q \bar{q} ) + [q \bar{Q}](Q \bar{q} ) +
[q \bar{Q}](\bar{q} Q  )
\eea
and thus $\eta = 1$ in Equs.\ \ref{equ:PVHVP} and \ref{equ:PVVP}.

It is sufficient to consider one ordering as long as we are consistent in our conventions.  We choose $A(\bar{Q}_1 q_2) B(Q_3 \bar{q}_4) \rightarrow A'(\bar{Q}_1 q'_2) B'(Q_3 \bar{q'}_4)$ where $q$ and $\bar{q}$ are the light $u$ and $d$ quarks and antiquarks, and $Q$ and $\bar{Q}$ are the heavy quarks and antiquarks. In this convention the spin and isospin operators act on the light quarks/antiquarks $i=2$ and $j=4$.

For simplicity of presentation just consider the terms where the heavy $\bar{Q}$ and $Q$ are in positions 1 and 3 respectively; the analysis trivially applies to all other combinations with the same conclusion.  The relevant terms are then

\be
|P\bar{V} \rangle = (- \bar{Q}^\uparrow q^\downarrow Q^\uparrow \bar{q}^\uparrow + \bar{Q}^\downarrow q^\uparrow Q^\uparrow \bar{q}^\uparrow)
\label{pvbar}
\ee

and

\be
|V\bar{P} \rangle = (- \bar{Q}^\uparrow q^\uparrow Q^\uparrow \bar{q}^\downarrow + \bar{Q}^\uparrow q^\uparrow Q^\downarrow \bar{q}^\uparrow)
\label{vpbar}
\ee

The $\pi$ exchange leaves the spins of the $Q (\bar{Q})$ unchanged and the $V \leftrightarrow P$ spin transition comes from the $q (\bar{q})$.
The non-zero transitions are then between the first term in Equ.\ \ref{pvbar} and the first term in Equ.\ \ref{vpbar}.
In each case the operator $\sigma_- \sigma_+$ gives $+1$ and hence the overall sign from the spin contributions to
$\langle P\bar{V} | H | V \bar{P} \rangle = +$

Hence the sign is

\bea
\langle P\bar{V} \pm \bar{P} V  | H |P\bar{V} \pm \bar{P} V  \rangle 
= \langle P\bar{V} \pm \bar{P} V  | H |\bar{V}P \pm V \bar{P}   \rangle
= \pm \equiv {Sign}(C)
\label{Equ:signpv}
\eea

The case of $V\bar{V}$ with all spins aligned, $S=S_z=2$, in the conventions above is

\bea
|V\bar{V} \rangle = [\bar{Q} q][Q\bar{q} ] - [\bar{Q} q][\bar{q} Q ] - [q \bar{Q} ][Q \bar{q} ] + [q \bar{Q} ][\bar{q} Q]
\eea
Here again, focusing on the terms where the heavy $\bar{Q}$ and $Q$ are in positions 1 and 3 respectively, and noting that the
$Q(\bar{Q})$ spins do not flip, it is immediately obvious that the sign is positive:

\bea
\langle V\bar{V} | H |V\bar{V}  \rangle = +
\label{Equ:signvv}
\eea
and hence the same as that for the $(P\bar{V} + \bar{P} V)$, $C=+$ channel $1^{++}$.

The explicit inclusion of flavour (isospin) for the $q\bar{q} \equiv d\bar{d} \pm u\bar{u}$ (and appropriate charge conjugated form)
introduces further signs, causing $I=0$ and $I=1$ channels to have opposite behaviours. However these factors are common to
all of the above and do not change the general conclusion that for a given isospin 
the $C=+$ combination of $(P\bar{V} + V\bar{P})$ has the {\it same}
sign as does $V\bar{V}$ in $J_z = 2$.

It is possible to deduce the overall sign as follows.

Start from $NN$ with $S=1$ and $I=0$: the deuteron.  This involves $\pi$ exchange between $q$ and $q$; there is no $q - \bar{q}$ interaction here.  Now consider the case of $VV$ with $S=2$ and $I=0$, which is like the deuteron in that again there is no $q - \bar{q}$ interaction.  The $\langle \vect{\sigma_i} \cdot \vect{\sigma_j} \rangle$ has the same sign in both cases.  The $\pi$ exchange (G-parity) gives opposite sign between $q-q$ and $q-\bar{q}$, and hence $V\bar{V}$ with $S=2$ and $I=0$ has opposite overall sign.  So far everything agrees with the calculations in Ref.\ \cite{Tornqvist:1993ng}.  It is only now, where Equs.\ \ref{Equ:signpv} and \ref{Equ:signvv} imply that the $C=+$ combination of $(P\bar{V} + V\bar{P})$ has the {\it same} sign as does $V\bar{V}$ in $^5S_2$, and hence opposite to the deuteron, in contrast to Ref.\ \cite{Tornqvist:1993ng}.

We shall show that when the spatial matrix elements are calculated, we agree with the formulation in Refs.\ \cite{Suzuki:2005ha,Liu:2008fh,Liu:2008du} and disagree with \cite{Tornqvist:1993ng}. This ironically introduces a further relative sign in the spatial contribution of Ref.\ \cite{Tornqvist:1993ng} in the charm sector, which, as we shall show, will eventually cause the $D\bar{D}^* + \bar{D} D^*$ state to be mildly attractive, in part as a result of two sign errors mutually cancelling.  However, the spatial sign-flip does not occur in the heavy quark limit, and hence some care is required in comparing $B\bar{B}^* + \bar{B} B^*$ to $D\bar{D}^* + \bar{D}D^*$.

\subsection{Overall Sign}
\label{sec:OverallSign}

The overall sign is determined, inter alia, by the expectation values of $(\vect{\sigma}_i \cdot \vect{\sigma}_j)(\vect{\tau}_i \cdot \vect{\tau}_j)$ and $S_{ij}(\vect{\hat{q}})(\vect{\tau}_i \cdot \vect{\tau}_j)$.  

The spin matrix element of the central term, $(\vect{\sigma}_i \cdot \vect{\sigma}_j)$, can be calculated explicitly as above or using general angular momentum theory.  General expressions are given in Appendix \ref{sec:SpinMatrixElements}.  It is straightforward to show that the spin matrix elements for $PP \rightarrow PP$, $PV \rightarrow PV$ and $VP \rightarrow VP$ vanish.  For $PV \rightarrow VP$ and $VP \rightarrow PV$ we get $+1$ and for $VV \rightarrow VV$ we get $\frac{1}{2}(S_T(S_T+1)-4)$.  These results agree with the calculation given above (Section \ref{sec:QuarkStates}).

The spin matrix element of the tensor term, $S_{ij}(\vect{\hat{q}}) \equiv 3(\vect{\sigma}_i \cdot \vect{\hat{q}})(\vect{\sigma}_j \cdot  \vect{\hat{q}}) - (\vect{\sigma}_i \cdot \vect{\sigma}_j)$, can be calculated in a similar way, see Appendix \ref{sec:SpinMatrixElements}.

The isospin factor is trivial to evaluate:
\begin{equation}
(\vect{\tau}_i \cdot \vect{\tau}_j) = \frac{1}{2}\left( (\vect{\tau}_i+\vect{\tau}_j)^2 - \vect{\tau_i}^2 - \vect{\tau_j}^2 \right) .
\end{equation}
For two isospin half mesons interacting, this is $2I(I+1)-3$, i.e. $-3$ in total isospin $I=0$ or $+1$ in $I=1$.

The interaction potential between a quark and an antiquark is opposite to that between two quarks (or two antiquarks) because of 
the G-parity of the pion.

Adopting the above results, for total spin $S$, isospin states with isospin $I$ and charge conjugation parity $C$, the flavour and spin factors for the central term are:
\begin{itemize}
\item $VV$: $(S(S+1)-4)(I(I+1)-3/2)$
\item $V\bar{V}$: $-(S(S+1)-4)(I(I+1)-3/2)$
\item $PV$: $(2I(I+1)-3)$
\item $P\bar{V}$: $-C(2I(I+1)-3)$
\end{itemize}

The $VV$ and $V\bar{V}$ expressions agree with those of Tornqvist\cite{Tornqvist:1993ng}.  However, there are different overall minus signs in the $PV$ and $P\bar{V}$ expressions compared to those of Tornqvist.

For reference we note that the matrix elements for $D \bar{D}^*$ with $J^{P}=1^{+}$, charge conjugation parity $C$ and isospin $I$, in the basis $L=0$, $L=2$ are \begin{equation}
 -C \left\{2I(I+1)-3\right\} \left[ 
\left( \begin{array}{cc} 1 & 0 \\ 0 & 1 \end{array} \right) V_C + 
\left( \begin{array}{cc} 0 & -\sqrt{2} \\ -\sqrt{2} & 1 \end{array} \right) V_T
\right]
\end{equation}
where $V_C$ and $V_T$ distinguish the central and tensor terms.

When isospin symmetry is broken, we replace the $(\vect{\tau}_1 \cdot \vect{\tau}_2)$ factor by a 2 by 2 matrix 
$$ \left( \begin{array}{cc} -1 & -2 \\ -2 & -1 \end{array} \right)\ $$ in the basis of charged/neutral states $00, +-$.  In the isospin limit, it is easy to see that this has eigenvalues $-3$ and $1$ with the respective isospin eigenvectors.  Close and Page\cite{Close:2003sg} and Tornqvist\cite{Tornqvist:2004qy} discuss isospin symmetry breaking. 

If one only considers the particular interaction $D^0 \bar{D}^{*0} \rightarrow D^{*0} \bar{D}^{0}$ and not every charge combination, the interaction strength is reduced by a factor of $1/3$ compared to the isospin symmetry limit with $I=0$.  Liu et.\ al.\cite{Liu:2008fh} only consider this particular charge interaction.

We can not really say whether the potentials are attractive or repulsive until we know something about the kinematic dependence which we discuss in the next section.

\section{Effective Potential in Position Space}
\label{sec:PositionSpacePotential}

If the interaction between $\pi$ and a light quark is taken to be
\begin{equation}
\mathcal{L} = \frac{g}{f_{\pi}} \bar{q}(x) \gamma^{\mu} \gamma_5 \vect{\vect{\tau}} q(x) \cdot \partial_{\mu} \vect{\phi}(x)  ,
\end{equation}
the effective potential between two light quarks due to one pion exchange in the static limit is
\begin{equation}
V(\vect{q}) =  \frac{g^2}{f_{\pi}^2}\frac{(\vect{\sigma}_i \cdot \vect{q})(\vect{\sigma}_j \cdot \vect{q})}{q^2 - m_{\pi}^2} (\vect{\tau}_i 
\cdot \vect{\tau}_j).
\end{equation}
Here $m_{\pi}$ is the $\pi$ mass, $f_{\pi}$ the $\pi$ decay constant (defined by $\sqrt{2}\left<0|A_{\mu}(0)|\pi^0(q)\right> = i f_{\pi^0} q_{\mu}$ or $\left<0|A_{\mu}(0)|\pi^{\pm}(q)\right> = i f_{\pi^{\pm}} q_{\mu}$ as in the PDG Review 2006\cite{PDG06}), $g$ a dimensionless coupling constant, and $q$ is the four-momentum 
transfer.  For elastic scattering this reduces to the form given in Ericson and 
Weise\cite{EricsonWeise:Pions}:
\begin{equation}
V(\vect{q}) = - \frac{g^2}{f_{\pi}^2}\frac{(\vect{\sigma}_i \cdot \vect{q})(\vect{\sigma}_j \cdot \vect{q})}{\mq^2 +
 m_{\pi}^2}(\vect{\tau}_i \cdot \vect{\tau}_j)
\end{equation}
\begin{equation}
V(\vect{r}) = \frac{g^2}{f_{\pi}^2}(\vect{\tau}_i \cdot \vect{\tau}_j)(\vect{\sigma}_i \cdot \grad)(\vect{\sigma}_j \cdot 
\grad)\frac{e^{-m_{\pi} r}}{4 \pi r} 
\end{equation}
where $\vect{q}$ is the three-momentum transfer.

Following Tornqvist\cite{Tornqvist:1993ng} we define $\mu^2 \equiv m_{\pi}^2 - (m_V-m_P)^2$, $m_V$ is the vector meson mass and $m_P$ the pseudoscalar meson mass.  Close to the static limit (both initial and final 3-momenta are zero) we have $q_0^2 \approx (m_V-m_P)^2$ and so $q^2 - m_{\pi}^2 \approx - \mq^2 - \mu^2$.  In the limit of elastic scattering $\mu = m_{\pi}$.  

We write the overall scale in terms of $V_0$:
\begin{equation}
V_0 \equiv \frac{m_{\pi}^3 g^2}{12 \pi f_{\pi}^2} 
\end{equation}
and so the expression for the effective potential is
\begin{equation}
V(\vect{q}) = - \frac{12 \pi V_0}{m_{\pi}^3} \frac{(\vect{\sigma}_i \cdot \vect{q})(\vect{\sigma}_j \cdot \vect{q})}
{\mq^2 + \mu^2}(\vect{\tau}_i \cdot \vect{\tau}_j)  .
\end{equation}
This is the same as Tornqvist's except for an overall minus sign -- this missing sign in his momentum space form
turns out not to be important because
 his expression for the potential in position space does have the correct sign.

Note that there are other definitions of $\mu$ in the literature.  Suzuki\cite{Suzuki:2005ha} and Liu et.\ al.\cite{Liu:2008fh}
 use $2(m_V - m_P - m_{\pi})m_{\pi} \approx - \mu^2$, where the correspondence, with a minus sign, is exact in the static limit. 
  As mentioned by Suzuki, the form of the potential in position space depends on the sign of $\mu^2$.

Following Ericson and Weise\cite{EricsonWeise:Pions}, it is useful to highlight the different spin dependences in the potential. 
 The central potential is proportional to $(\vect{\sigma}_i \cdot \vect{\sigma}_j)$ (hyperfine interaction-like) and the tensor term is
  proportional to $S_{ij}(\vect{\hat{q}})$:
\begin{equation}
\label{Equ:Vq}
V(\vect{q}) = \frac{4 \pi V_0}{m_{\pi}^3} \left[ \left(-1 + \frac{\mu^2}{\mq^2+\mu^2} \right) (\vect{\sigma}_i \cdot \vect{\sigma}_j)
 - \frac{\mq^2}{\mq^2+\mu^2} S_{ij}(\vect{\hat{q}}) \right] (\vect{\tau}_i \cdot \vect{\tau}_j)
\end{equation}
with $S_{ij}(\vect{\hat{q}}) \equiv 3(\vect{\sigma}_i \cdot \vect{\hat{q}})(\vect{\sigma}_j \cdot \vect{\hat{q}}) - (\vect{\sigma}_i 
\cdot \vect{\sigma}_j)$.  The literature\cite{EricsonWeise:Pions,Tornqvist:1991ks,Tornqvist:1993ng,Tornqvist:2004qy,Swanson:2003tb,Suzuki:2005ha,Liu:2008fh} has focused on the potential in position space but disagree on its form. Hence we take the Fourier transform and compare with the literature.

For $\mu^2>0$:
\begin{equation}
V(\vect{r}) = V_0 \left[ \left(-\frac{4\pi}{m_{\pi}^3} \delta(\vect{r}) + \frac{\mu^2}{m_{\pi}^2} \frac{e^{-\mu r}}{m_{\pi} r} \right) 
(\vect{\sigma}_i \cdot \vect{\sigma}_j) + \frac{\mu^2}{m_{\pi}^2} \frac{e^{-\mu r}}{m_{\pi} r}\left(1+\frac{3}{\mu r}+\frac{3}{\mu^2 r^2} 
\right)S_{ij}(\vect{\hat{r}}) \right] (\vect{\tau}_i \cdot \vect{\tau}_j) 
\end{equation}
and for $\mu^2 \equiv -\tmu^2 < 0$ the real part is
\begin{equation}
V(\vect{r}) = V_0 \left[ \left(-\frac{4\pi}{m_{\pi}^3} \delta(\vect{r}) - \frac{\tmu^2}{m_{\pi}^2} \frac{\cos{\tmu r}}{m_{\pi} r} \right) 
(\vect{\sigma}_i \cdot \vect{\sigma}_j) - \frac{\tmu^2}{m_{\pi}^2} \frac{1}{m_{\pi} r}\left(\cos{\tmu r}-\frac{3\sin{\tmu r}}{\tmu r}
-\frac{3\cos{\tmu r}}{\tmu^2 r^2} \right)S_{ij}(\vect{\hat{r}}) \right] (\vect{\tau}_i \cdot \vect{\tau}_j)  
\end{equation}
where we have kept the terms in the same order.

The expression with $\mu^2>0$ reproduces that of Tornqvist\cite{Tornqvist:1993ng}, except that he ignores the $\delta(\vect{r})$ term.  
He allows for a varying $\mu^2$ but his quoted potential appears implicitly to assume that $\mu^2$ is always positive.  
In cases where $\mu^2<0$, our potential agrees with that of 
Liu et.\ al.\cite{Liu:2008fh} (except that they have ignored the tensor term)
and with that of Suzuki\cite{Suzuki:2005ha} apart from some relative minus signs (where we confirm the form of Liu et.\ al.).

\subsubsection*{Regularization}

The potential is singular at small distances and so has to be regularised.  Following Tornqvist and Liu et. al., we do this by introducing a form factor at each $\pi$ vertex which leads to an extra factor of $\left(\frac{\Lambda^2 - m_{\pi}^2}{\Lambda^2 - q^2}\right)^2$ multiplying $V(\vect{q})$ (Equ.\ \ref{Equ:Vq}).  This gives the $\pi$ an effective RMS radius of $\sqrt{10}/\Lambda$\cite{Tornqvist:1993ng}.  The value of $\Lambda$ has to be determined phenomenologically.  Tornqvist mentions that in nucleon-nucleon interactions values between $0.8$ and $1.5\ \GeV$ have been used depending on the model and application, but that larger values ($\Lambda > 1.4\ \GeV$) are required for nucleon-nucleon phase shifts.  He says that for heavy mesons which have a smaller size than nucleons, one would expect a smaller effective radius of the $\pi$ source corresponding to a larger $\Lambda$.  A larger $\Lambda$ gives a stronger potential at short distances; we shall find in Section \ref{sec:Applications} that the results depend strongly on $\Lambda$.

The central terms (i.e. all but the tensor $S_{ij}(\vect{\hat{r}})$) for $\mu^2>0$ are
\begin{equation}
\label{Equ:FormFact:spin:mup}
V_0 \left[ -\frac{X(\Lambda^2 - m_{\pi}^2)}{2 m_{\pi}^3}e^{-X r} + \frac{\mu^2}{m_{\pi}^3 r} \left(e^{-\mu r} - e^{-X r} \right)
 \right] (\vect{\sigma}_i \cdot \vect{\sigma}_j)(\vect{\tau}_i \cdot \vect{\tau}_j) 
\end{equation}
with $X^2 \equiv \Lambda^2 + \mu^2 - m_{\pi}^2$.  Tornqvist omits the `$\delta$ function' piece (more precisely the piece that gave the delta function when no 
regularisation was used), ``which from the phenomenological point of view will be included in the short range potential and 
regularisation scheme''\cite{Tornqvist:1993ng}.  This results in all his terms having a common $\mu^2$ factor.  
If for comparison we also ignore the `$\delta$ function' piece we get
\begin{equation}
\label{Equ:FormFact:spin:nodeltafn:mup}
V_0 \left[ -\frac{1}{2}\frac{\mu^2(\Lambda^2 - m_{\pi}^2)}{X m_{\pi}^3}e^{-X r} + \frac{\mu^2}{m_{\pi}^3 r} \left(e^{-\mu r} - e^{-X r} 
\right) \right] (\vect{\sigma}_i \cdot \vect{\sigma}_j)(\vect{\tau}_i \cdot \vect{\tau}_j) 
\end{equation}
Note that this is not quite the same as the expression in Ref.\ \cite{Tornqvist:1993ng}, namely:
\begin{equation}
\label{Equ:FormFact:spin:Tornqvist}
V_0 \left[ -\frac{\mu^2(\Lambda^2 - \mu^2)}{\Lambda m_{\pi}^3}e^{-\Lambda r} + \frac{\mu^2}{m_{\pi}^3 r} \left(e^{-\mu r} - e^{-\Lambda r} \right) \right] (\vect{\sigma}_i \cdot \vect{\sigma}_j)(\vect{\tau}_i \cdot \vect{\tau}_j)  .
\end{equation}
The most significant difference (e.g. see Figures \ref{fig:CentralTerms_Mup} and \ref{fig:CentralTerms_Mun}) is due to the missing factor of $1/2$ in the first term.  The other differences are a $m_{\pi}^2 \rightarrow \mu^2$ in one place and the approximation $X \approx \Lambda$.

We now comment on the case $\mu^2 \equiv -\tmu^2 < 0$. The real part of the full central term (including the `$\delta$ function' piece) is
\begin{equation}
\label{Equ:FormFact:spin:mun}
V_0 \left[ -\frac{X(\Lambda^2 - m_{\pi}^2)}{2 m_{\pi}^3}e^{-X r} -\frac{\tmu^2}{m_{\pi}^3 r} \left(\cos(\tmu r) - e^{-X r} \right)
 \right] (\vect{\sigma}_i \cdot \vect{\sigma}_j)(\vect{\tau}_i \cdot \vect{\tau}_j)  .
\end{equation}
This expression agrees with Equ.\ 19 of Liu et.\ al.  If we ignore the `$\delta$ function' piece we get
\begin{equation}
\label{Equ:FormFact:spin:nodeltafn:mun}
V_0 \left[ +\frac{\tmu^2(\Lambda^2 - m_{\pi}^2)}{2 X m_{\pi}^3}e^{-X r} -\frac{\tmu^2}{m_{\pi}^3 r} \left(\cos(\tmu r) - e^{-X r} 
\right) \right] (\vect{\sigma}_i \cdot \vect{\sigma}_j)(\vect{\tau}_i \cdot \vect{\tau}_j)  .
\end{equation}

The tensor terms for $\mu^2>0$ are
\begin{eqnarray}
\label{Equ:FormFact:tensor:mup}
V(\vect{r}) &=& V_0 \left[ \frac{\mu^2}{m_{\pi}^2} \frac{e^{-\mu r}}{m_{\pi} r}\left(1+\frac{3}{\mu r}+\frac{3}{\mu^2 r^2} \right) 
\right. \\
\nonumber && \left. - \frac{X^2}{m_{\pi}^2} \frac{e^{-X r}}{m_{\pi} r}\left(1+\frac{3}{X r}+\frac{3}{X^2 r^2} \right) - 
\frac{(\Lambda^2 - m_{\pi}^2)}{2 m_{\pi}^2}\frac{e^{-X r}}{m_{\pi} r}(1+Xr) \right] S_{ij}(\vect{\hat{r}}) (\vect{\tau}_i
 \cdot \vect{\tau}_j) .
\end{eqnarray}
This agrees with the Equ.\ 26 of Tornqvist apart from one place in the last term where $m_{\pi}^2 \rightarrow \mu^2$.
For $\mu^2 \equiv -\tmu^2 < 0$ the real part of the tensor terms are
\begin{eqnarray}
\label{Equ:FormFact:tensor:mun}
V(\vect{r}) &=& V_0 \left[ -\frac{\tmu^2}{m_{\pi}^2} \frac{1}{m_{\pi} r}\left(\cos{\tmu r}-\frac{3\sin{\tmu r}}{\tmu r}-
\frac{3\cos{\tmu r}}{\tmu^2 r^2} \right) \right. \\
\nonumber && \left. - \frac{X^2}{m_{\pi}^2} \frac{e^{-X r}}{m_{\pi} r}\left(1+\frac{3}{X r}+\frac{3}{X^2 r^2} \right) - 
\frac{(\Lambda^2 - m_{\pi}^2)}{2 m_{\pi}^2}\frac{e^{-X r}}{m_{\pi} r}(1+Xr) \right] S_{ij}(\vect{\hat{r}}) (\vect{\tau}_i 
\cdot \vect{\tau}_j) .
\end{eqnarray}

We collate some useful relationships and Fourier transforms in Appendix \ref{sec:PositionSpaceExpressions} from which the derivation of the above expressions may be checked.

We shall use the forms Equs.\ \ref{Equ:FormFact:spin:mup} and \ref{Equ:FormFact:spin:mun} for the central term, and Equs.\ \ref{Equ:FormFact:tensor:mup} and \ref{Equ:FormFact:tensor:mun} for the tensor term in our analysis.  We shall illustrate the implications for binding by comparing with other expressions in the previous literature.

\section{Applications}
\label{sec:Applications}

\subsection{Normalisation}
\label{sec:Normalisation}

We have defined $V_0$ in the same way as Tornqvist\cite{Tornqvist:1993ng} and so we can write it in terms of the $\pi$-nucleon 
coupling constant $f_{\pi N}$:
\begin{equation}
\frac{f_{\pi N}^2}{4\pi} = \frac{25}{9} \frac{g^2 m_{\pi}^2}{f^2 4\pi} = \frac{25}{9} \frac{3 V_0}{m_{\pi}} \approx 0.08  .
\end{equation}
Tornqvist finds $V_0 \approx 1.3\ \MeV$, which is consistent with Ericson and Weise\cite{EricsonWeise:Pions} and Ericson and 
Karl\cite{Ericson:1993wy}.  

$V_0$ can also be related to the $D^* \rightarrow D^0 \pi^+$ transition:
\begin{equation}
\Gamma(D^* \rightarrow D^0 \pi^+) = \frac{g^2}{6 \pi f_{\pi}^2} p_{\pi}^3 = 2 V_0 \frac{p_{\pi}^3}{m_{\pi}^3} .
\end{equation}
The PDG Review 2006\cite{PDG06} gives $\Gamma(D^* \rightarrow D^0 \pi^+) = (65 \pm 15) \keV$ and so $V_0 = (1.5 \pm 0.3)\ \MeV$ 
which is consistent with the above.

Suzuki\cite{Suzuki:2005ha} appears to have an effective $V_0 \approx 0.73\ \MeV$ and Liu et.\ al.\cite{Liu:2008fh} have $V_0 \approx 0.68\ \MeV$.  Note that there are various definitions of the $\pi$ decay constant used in the literature.  We take the definition of the PDG Review 2006\cite{PDG06} where $f_{\pi}\approx130\ \MeV$.  Compare this with the definition used in Swanson\cite{Swanson:2003tb} where $\frac{g^2}{f_{\pi}^2} \rightarrow \frac{g^2}{2f_{\pi}^2}$ and $f_{\pi} = 92\ \MeV$.

We will take $V_0 = 1.3\ \MeV$ throughout this work.

\subsection{Shape of The Potentials}
\label{sec:PlotPotentials}

\begin{figure}[htb]
\begin{center}
\includegraphics[width=15cm]{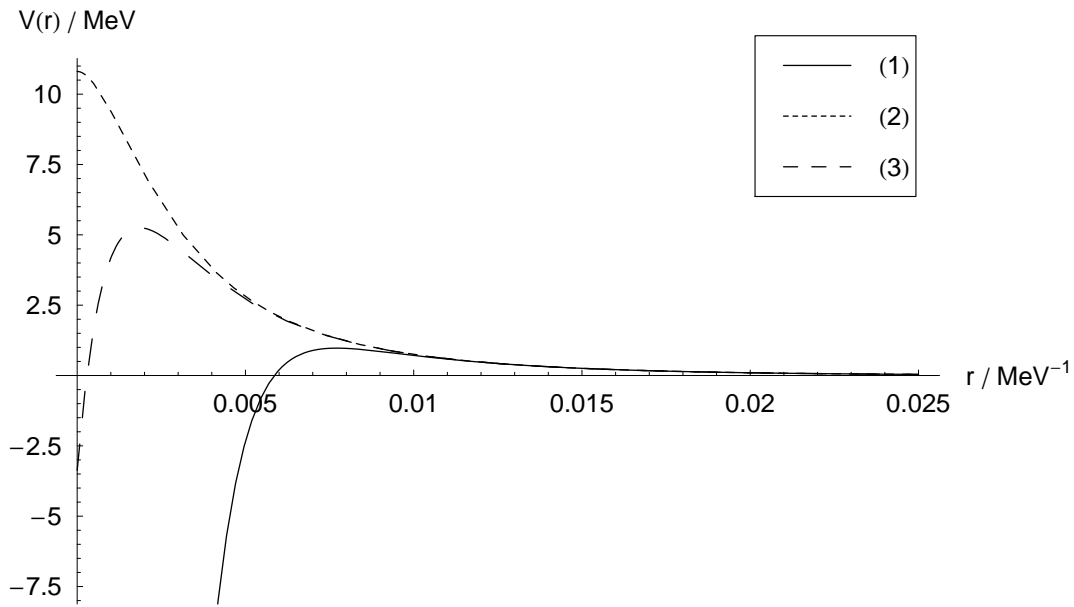}
\caption{Central potential for $P\bar{V}$ in $I=0$, $L=0$, $J^{PC}=1^{++}$ with $\Lambda = 1\ \GeV$, $V_0 = 1.3\ \MeV$, and $\mu^2 = m_{\pi}^2$.  (1) is our potential (Equ.\ \ref{Equ:FormFact:spin:mup}), (2) is our potential without the `$\delta$ function' piece (Equ.\ \ref{Equ:FormFact:spin:nodeltafn:mup}),  to be compared with Tornqvist's analogous (i.e. no $\delta$-function) potential (Equ.\ \ref{Equ:FormFact:spin:Tornqvist}) shown as curve (3).}
\label{fig:CentralTerms_Mup}
\end{center}
\end{figure}

\begin{figure}[htb]
\begin{center}
\includegraphics[width=15cm]{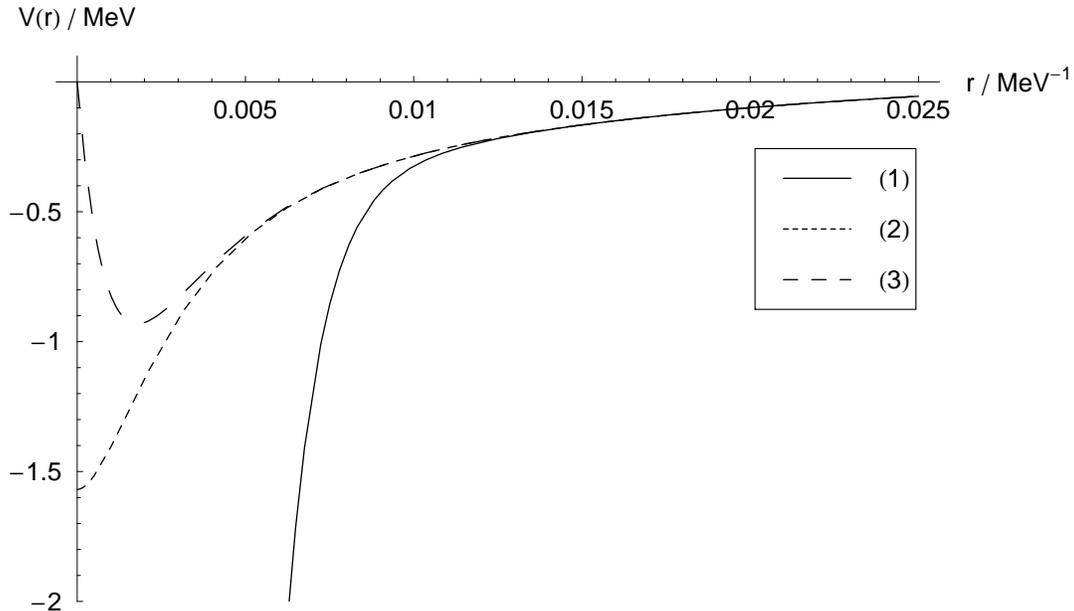}
\caption{Central potential for $P\bar{V}$ in $I=0$, $L=0$, $J^{PC}=1^{++}$ with $\Lambda = 1\ \GeV$, $V_0 = 1.3\ \MeV$, and $\mu^2 = -2000\ \MeV^2$.  (1) is our potential (Equ.\ \ref{Equ:FormFact:spin:mun}), (2) is our potential without the `$\delta$ function' piece (Equ. \ref{Equ:FormFact:spin:nodeltafn:mun}), (3) is Tornqvist's potential modified for negative $\mu^2$.}
\label{fig:CentralTerms_Mun}
\end{center}
\end{figure}

In Figures \ref{fig:CentralTerms_Mup} and \ref{fig:CentralTerms_Mun} we plot the central part of the potential $P\bar{V}$ in $I=0$, $L=0$, $J^{PC}=1^{++}$ with $\Lambda = 1\ \GeV$, $V_0 = 1.3\ \MeV$, and $\mu^2 = m_{\pi}^2$ or $\mu^2 = -2000\ \MeV^2$ respectively.  As well as our expression for the full central potential (Equs.\ \ref{Equ:FormFact:spin:mup} and \ref{Equ:FormFact:spin:mun}), for comparison we plot the potential without the `$\delta$ function' piece (Equs. \ref{Equ:FormFact:spin:nodeltafn:mup} and \ref{Equ:FormFact:spin:nodeltafn:mun}) and Tornqvist's expression for the potential (Equ. \ref{Equ:FormFact:spin:Tornqvist}).  Note that in all three cases we use our expression for the spin and flavour factors as discussed in Section \ref{sec:OverallSign}.

Figure \ref{fig:CentralTerms_Mup} shows that for positive $\mu^2$ the `$\delta$ function' term is attractive, dominates at 
short distances and is \emph{opposite} in sign to the other central terms.  The effect of the different factor of $2$ 
in Tornqvist's potential compared to ours when we ignore the `$\delta$ function' term can also be seen, in particular, the sign change as $r \rightarrow 0$.  The three potentials differ at short distances but become indistinguishable for $r \gtrsim 2\ \text{fm}$.  

Figure \ref{fig:CentralTerms_Mun} shows that for negative $\mu^2$ the `$\delta$ function' term remains attractive, dominates at short distances and is the \emph{same sign} as the other central terms.  The effect of the different factor of $2$ in Tornqvist's potential can again be seen, as can the fact that the three potentials differ at short distances but are the same at long distances.  As is apparent from comparing Figures \ref{fig:CentralTerms_Mup} and \ref{fig:CentralTerms_Mun}, the `$\delta$ function' term does not change sign when $\mu^2 \rightarrow -\mu^2$ whereas the other central terms do.  Thus the phenomenological conclusions may depend significantly on how the $\delta$-function term is treated.

\begin{figure}[htb]
\begin{center}
\includegraphics[width=15cm]{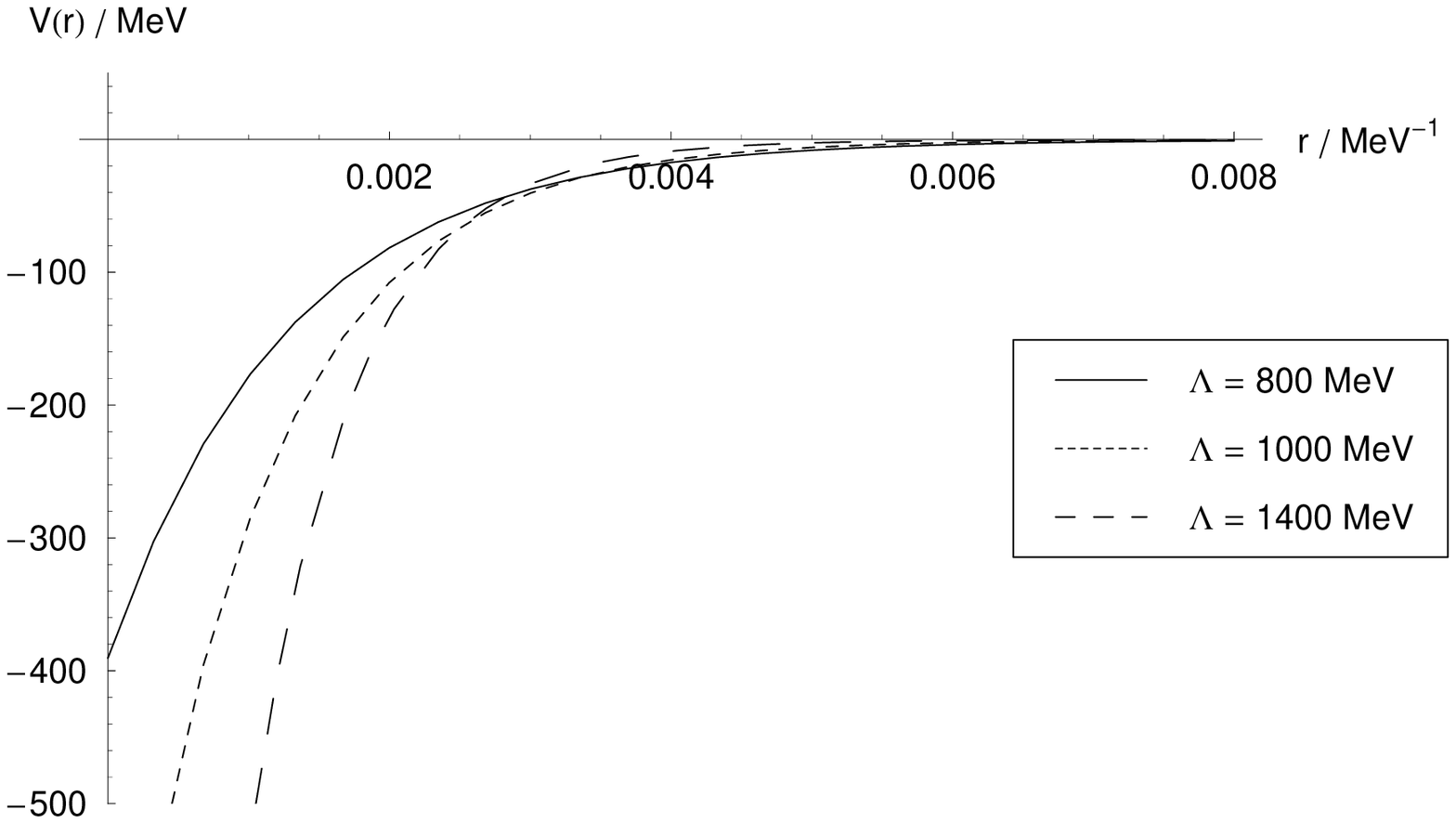}
\caption{Central potential for $P\bar{V}$ in $I=0$, $L=0$, $J^{PC}=1^{++}$ with $V_0 = 1.3\ \MeV$, $\mu^2 = -2000\ \MeV^2$, and with our expression for the potential (Equ.\ \ref{Equ:FormFact:spin:mun}) for various values of $\Lambda$.}
\label{fig:CentralTerms_Gamma}
\end{center}
\end{figure}

In Figure \ref{fig:CentralTerms_Gamma} we plot our expression for the central potential with the above parameters, $\mu^2 = -2000\ \MeV^2$ and show the effect of varying $\Lambda$.  It can be seen that increasing $\Lambda$ increases the strength of the potential at short distances and that the potential is quite sensitive to $\Lambda$.

\subsection{Application to $D \bar{D}^*$}
\label{sec:DDv}

In this section we discuss application of the above formalism to the $D \bar{D}^*$ system with $J^{PC} = 1^{++}$.  The mesons can be in two different partial waves ($L=0$ or $2$) and mixing between these may be important.

We solve the potential by discretising the time-independent Schrodinger equation with the one pion exchange potential and then diagonalising the resulting matrix.  We vary the number of points and the maximum radius to check for discretisation and finite volume effects.

To get some handle on the size of $\Lambda$, we first calculate and solve the effective potential for the deuteron.  We assume isospin symmetry and include both the central and tensor terms.  For comparison, we consider three different expressions for the central potential: (1) our full expression (Equs.\ \ref{Equ:FormFact:spin:mup} and \ref{Equ:FormFact:spin:mun}), (2) Our expression without the `$\delta$ function' piece (Equ.\ \ref{Equ:FormFact:spin:nodeltafn:mup} and \ref{Equ:FormFact:spin:nodeltafn:mun}), and (3) Tornqvist's expression (Equ.\ \ref{Equ:FormFact:spin:Tornqvist}).  We vary $\Lambda$ to fix the binding energy to $\approx 2.2\ \MeV$.

For potentials (1), (2) and (3) we find the values required are $\Lambda = 960\ \MeV$, $750\ \MeV$ and $760\ \MeV$ respectively.  More details of the calculations and results are given in Appendix \ref{sec:Deuteron}.  $\Lambda$ gives an effective radius of the pion $\sim 1/\Lambda$ which is expected\cite{Tornqvist:1993ng} to depend on the interacting hadron and so be smaller for the $D \bar{D}^*$ system compared to the deuteron.  This argument implies that the $\Lambda$ we have found from solving the deuteron is a lower bound on the value of $\Lambda$ that should be used for the $D \bar{D}^*$ system.  The value of $\Lambda$ required for the $B \bar{B}^*$ system should be larger again.

Moving to the $D \bar{D}^*$ system, we verify the results of Ref.\ \cite{Liu:2008fh} within the assumptions made there (noting the possible difference in normalisation discussed above in Section \ref{sec:Normalisation}).  Specifically, we initially considered only the $D^0 \bar{D}^{*0} + \bar{D}^0 D^{*0}$ charge combination, used our full expression for the central term (1) and ignored the tensor term.  In this case the flavour factor is $-1$ and one pion exchange is slightly inelastic with a negative $\mu^2$ ($\mu^2 = -2000\ \MeV^2$).  For a reasonable range of $\Lambda$ (we checked up to $2.5\ \GeV$) we find no bound state, supporting the results of Ref.\ \cite{Liu:2008fh} within their assumptions.  However, for a larger $\Lambda$ of $3\ \GeV$, we do find a bound state with energy $\approx 3868\ \MeV$.

However, this shows how sensitive the results are to such assumptions.  For example, if we now assume isospin symmetry with $I=0$ (flavour factor is $-3$) and don't change anything else, we \emph{do} find a bound state.  For $\Lambda$ in the range $960\ \MeV$ to $1200\ \MeV$ we find $E = 3871\ \MeV - 3863 \MeV$.  This highlights the importance of including all the possible charge combinations.

We then performed a more complete calculation including the tensor term.  We also allow for isospin symmetry breaking through different charged/neutral meson masses and $\mu^2$s.  The $\mu^2$s are all \emph{negative} except for $D^{*0} \rightarrow D^+$ which has a \emph{positive} $\mu^2$.  Because we can only use one $\mu^2$ for each charge mode, there is ambiguity as to whether we should calculate the charged-neutral $\mu^2$ from (A) $D^{*+} \rightarrow D^0$ or (B) $D^{*0} \rightarrow D^+$.  We will compare the results obtained with both choices (A) and (B).

The detailed results are given in Table \ref{Table:DDvResults} of Appendix \ref{sec:ResultsTables}.  With our full expression for the potential (1), we find a bound state with $E = 3870\ \MeV - 3806\ \MeV$ for $\Lambda \approx 960\ \MeV - 1400\ \MeV$.  For $\Lambda = 800\ \MeV$ we find no bound state.  This shows that the binding energy is very sensitive to $\Lambda$.  The choice of $\mu^2$ (A) or (B) modifies the binding energy by at most $1\ \MeV$.  In particular, Table \ref{Table:DDvResults} shows that isospin symmetry breaking is important, especially when the state is close to threshold.  This is not particularly surprising: the charged-neutral mass difference gets relatively more important when the state is only just bound.  The state is mostly $L=0$ with a small $L=2$ component.

Using potentials (2) and (3), we require a larger $\Lambda$ for a bound state ($\Lambda \gtrsim 1750\ \MeV$) because these potentials do not have the strong attraction of the `$\delta$ function' term.  We find $E = 3871\ \MeV - 3863\ \MeV$ for $\Lambda \approx 1750\ \MeV - 2000\ \MeV$.  The differences between potentials (2) and (3) do not significantly change the binding energy or other properties, as shown in Table \ref{Table:DDvResults}.

In summary, we have found a bound state for reasonable values of $\Lambda$ using all three potentials.  The binding energy is sensitive to the potential and the value of $\Lambda$ used.  Clearly the exact values of the binding energies calculated are not significant, because of their sensitivity to parameters and the potential.  Nonetheless, the general conclusion that within one pion exchange potentials a bound state can be formed appears to be robust.  Conversely, there is no theoretical reason to expect that it must be formed, nor that it cannot occur.  Basically, the phenomena of the $X(3872)$, within the assumption of one $\pi$-exchange, would constrain the parameter $\Lambda$ more than present knowledge of this parameter can constrain the $DD^*$ dynamics.

\subsection{Further Applications}
\label{sec:FurtherApps}

We can also apply a similar analysis to the analogous $B \bar{B}^*$ system.  Furthermore isospin breaking is expected to be less important here because of the relatively smaller charged/neutral differences.  The larger $B$ meson masses mean that the kinetic energy is less of a hindrance to binding.

Results are given in Table \ref{Table:BBvResults} of Appendix \ref{sec:ResultsTables} which show that a bound state is found for reasonable values of $\Lambda$, the binding energy is sensitive to $\Lambda$ and isospin breaking is less important than in the $D\bar{D}^*$ system.  We note that there are larger $L=2$ components here compared to the $D \bar{D}^*$ system, especially when using potentials (2) and (3).  Because $\mu^2$ is positive here, the central terms other than the `$\delta$ function' term are repulsive and so the tensor term is important for forming a bound state.

The same analysis again can be applied to flavour exotics, that is states such as $D D^{*}$ and $B B^*$ with two charm or bottom quarks.  The one pion exchange interaction in such states has opposite overall sign to that in the $D \bar{D}^*$ and $B \bar{B}^*$ systems (Section \ref{sec:OverallSign}).  However, mixing with the tensor terms complicates the situation and we must do more than just consider the sign of the central term.

For the $D D^*$ system, results are given in Table \ref{Table:ExoticDDvResults} of Appendix \ref{sec:ResultsTables}.  With potential (1) we find a bound state with energy $\approx 3870\ \MeV$ for $\Lambda = 2000\ \MeV$.  There is no bound state for $\Lambda = 1700\MeV$.  The importance of the tensor term is apparent from the relatively large $L=2$ component.  The $+0$ and $0+$ components have the opposite sign and so the state would be isoscalar in the isospin symmetry limit. Using potential (2) or (3) we find a bound state with energy $\approx 3870\ \MeV$ for $\Lambda = 1500\ \MeV$, a bound state with energy $\approx 3810\ \MeV$ for $\Lambda = 2000\ \MeV$, and no bound state for $\Lambda = 1000\ \MeV$.  Again, there are relatively large $L=2$ components.

For the $B B^*$ system, results are given in Table \ref{Table:ExoticBBvResults} of Appendix \ref{sec:ResultsTables}.  With potential (1), we find a bound state with energy $\approx 10600\ \MeV$ with $\Lambda = 1000\ \MeV$, a bound state with energy $\approx 10560\ \MeV$ with $\Lambda = 1500\ \MeV$, and no bound state for $\Lambda = 700\ \MeV$.  There are again relatively large $L=2$ components and the $+0$ and $0+$ components have opposite sign.  Using potential (2) or (3) we find a bound state with energy $\approx 10600\ \MeV$ for $\Lambda = 700\ \MeV$, a bound state with energy $\approx 10590\ \MeV$ for $\Lambda = 1000\ \MeV$, and no bound state for $\Lambda = 500\ \MeV$.  Again there are relatively large $L=2$ components.

In summary, we find great sensitivity to parameters, but qualitatively confirm that $B\bar{B}^*$ and exotic states can bind in one pion exchange.  As can be seen from Table \ref{Table:LambdaSummary}, using our expression for the potential (1), we find $D\bar{D}^*$, $B\bar{B}^*$ and $BB^*$ can bind with $\Lambda \sim 1000\ \MeV$.  However, $D\bar{D}^*$ requires a larger $\Lambda \sim 2000\ \MeV$.  Hence, within our potential and assumptions, the parameters that allow $X(3872)$ to emerge as a bound state preclude binding the exotic $DD^*$ channel.  However, binding in both $B\bar{B}^*$ and exotic $BB^*$ are possible.

\begin{table}[htb]
\begin{center}
\begin{tabular}{|c|c|c|}
\hline
\textbf{System} & \textbf{Potential} & \textbf{Approximate $\bm{\Lambda / \MeV}$}  \\
\hline
$D\bar{D}^*$ & 1 & $960$  \\
 & 2 and 3 & $1750$  \\
\hline
$B\bar{B}^*$ & 1 & $<800$  \\
 & 2 and 3 & $1000$  \\
\hline
$DD^*$ & 1 & $2000$  \\
 & 2 and 3 & $1500$  \\
\hline
$BB^*$ & 1 & $1000$  \\
 & 2 and 3 & $700$  \\
\hline
\end{tabular}
\end{center}
\caption{Approximate minimum values of $\Lambda$ required to bind}
\label{Table:LambdaSummary}
\end{table}

\section{Conclusions}
\label{sec:Conclusions}

In summary, we agree with the qualitative results of Tornqvist, but as a result of various differences cancelling out.  Swanson\cite{Swanson:2003tb} found that quark exchange alone did not bind within the one gluon exchange contact approximation.  This led him to include one pion exchange based upon the work of Ref.\ \cite{Tornqvist:1991ks}.

We have quantified the arguments of Ref.\ \cite{Suzuki:2005ha} that a small $\mu^2$ leads to a small binding energy.  In doing so, we have included the tensor term, and also flavour factors, neither of which were discussed in that reference.  Our calculations show that the sensitivity to $\Lambda$ is the overriding factor.

Liu et.\ al.\cite{Liu:2008fh} considered only $D^0 \bar{D}^{0*}$; charged modes were ignored, as was the tensor term.  Within their assumptions we confirm their results, though there is the question of overall normalisation.  Our work highlights the importance of taking into account all charged modes.  They have also considered $\sigma$ exchange, and argue that this makes it harder for the $D \bar{D}^*$ to bind.

We have discussed flavour exotic states and find that the overall sign of the central term alone does not determine whether or not a bound state is formed;  mixing due to the tensor term is also important.  It can be dangerous to ignore the tensor term and its contribution to different processes can be important: this is especially true for the case of flavour exotics.  


\section*{Acknowledgements}

We are indebted to S.-L.\ Zhu, V.\ Lyubovitskij, E.\ Swanson, and N.\ Tornqvist for discussions of their work and to C.\ Downum, R.\ C.\ Johnson, Q.\ Zhao for other useful discussion.

This work is supported by a studentship and grants from the Science \& Technology Facilities Council (UK).


\appendix

\section{Spin Matrix Elements}
\label{sec:SpinMatrixElements}

We choose $A(\bar{Q}_1 q_2) B(Q_3 \bar{q}_4) \rightarrow A'(\bar{Q}_1 q'_2) B'(Q_3 \bar{q'}_4)$ where $q$ and $\bar{q}$ are the light $u$ and $d$ quarks and antiquarks, and $Q$ and $\bar{Q}$ are the heavy quarks and antiquarks.  In this convention the spin operators act on the light quarks/antiquarks $i=2$ and $j=4$.

In general, for initial state mesons with spin $S_A$ and $S_B$ coupled to total spin $S_T$ going  to final state mesons $S_A'$ and $S_B'$ coupled to total spin $S_T'$, the matrix element of the central term, $(\vect{\sigma}_2 \cdot \vect{\sigma}_4)$ is given by:
\begin{eqnarray}
&&\left< S_A' S_B' (S_T' M')|(\vect{\sigma}_2 \cdot \vect{\sigma}_4)|S_A S_B (S_T M) \right> = \\
\nonumber && \delta_{S_T',S_T} \delta_{M',M} \sum_{S_{13} S_{24}} \left[ 2\left(S_{24}(S_{24}+1)-3/2\right) 
\Pi_{S_A S_B S_A' S_B' S_{13} S_{13} S_{24} S_{24}}
\left\{ \begin{array}{ccc} 1/2 & 1/2 & S_A \\ 1/2 & 1/2 & S_B \\ S_{13} & S_{24} & S_T \end{array} \right\}
\left\{ \begin{array}{ccc} 1/2 & 1/2 & S_{13} \\ 1/2 & 1/2 & S_{24} \\ S_A' & S_B' & S_T' \end{array} \right\}
\right]
\end{eqnarray}
where $\{ \}$ are Wigner 9j symbols and $\Pi_{A B C ...} \equiv \sqrt{(2A+1)(2B+1)(2C+1)...}$ .

From this expression, it is straightforward to show that the spin matrix elements for $PP \rightarrow PP$, $PV \rightarrow PV$ and $VP \rightarrow VP$ vanish.  For $PV \rightarrow VP$ and $VP \rightarrow PV$ we get $+1$ and for $VV \rightarrow VV$ we get  $\frac{1}{2}(S_T(S_T+1)-4)$.  These results agree with the direct calculation given above.

The matrix element of the tensor term, $S_{24}(\vect{\hat{q}}) \equiv 3(\vect{\sigma}_2 \cdot \vect{\hat{q}})(\vect{\sigma}_4 \cdot
 \vect{\hat{q}}) - (\vect{\sigma}_2 \cdot \vect{\sigma}_4)$, can be calculated in a similar way\cite{Johnson:TensorMatrixElements}.  With the spins defined above and with the total spin $J_{AB}$, $J_{AB}'$, relative orbital angular momentum $L$, $L'$ and total angular momentum $J$, $J'=J$ in the initial and final states respectively, the matrix element is:
\begin{eqnarray}
&&\left<A' B'|S_{24}(\vect{\hat{q}})|A B\right> = \\
\nonumber && 4 \sqrt{\frac{5}{2}}\sum_{\tilde{J} S_{13} S_{24}} \left[ \delta_{S_{24},1} (-1)^{L'+J_{AB}+J_{AB}'+\tilde{J}+S_{24}}
\Pi_{S_A S_B S_A' S_B' S_{13} S_{13} S_{24} S_{24} S_{24} J_{AB} J_{AB}' \tilde{J} \tilde{J} L} \right. \\
\nonumber && \left.
\left\{ \begin{array}{ccc} 1/2 & 1/2 & S_A \\ 1/2 & 1/2 & S_B \\ S_{13} & S_{24} & J_{AB} \end{array} \right\}
\left\{ \begin{array}{ccc} 1/2 & 1/2 & S_A' \\ 1/2 & 1/2 & S_B' \\ S_{13} & S_{24} & J_{AB}' \end{array} \right\}
\left\{ \begin{array}{ccc} L & S_{24} & \tilde{J} \\ S_{13} & J & J_{AB} \end{array} \right\}
\left\{ \begin{array}{ccc} L' & S_{24} & \tilde{J} \\ S_{13} & J & J_{AB}' \end{array} \right\}
\left\{ \begin{array}{ccc} L' & L & 2 \\ S_{24} & S_{24} & \tilde{J} \end{array} \right\}
\CG{L}{2}{L'}{0}{0}{0}
\right]
\end{eqnarray}
where $\{ \}$ are Wigner 6j and 9j symbols and $\CG{L}{2}{L'}{0}{0}{0}$ is a Clebsch Gordan coefficient.

\section{Useful expressions for calculating the potential in position space}
\label{sec:PositionSpaceExpressions}

We collate useful Fourier transforms used in Table \ref{Table:FourierTransforms} where 
\begin{equation}
V(\vect{r}) = \frac{1}{(2\pi)^3} \int {V(\vect{q}) e^{i\vect{q}\cdot\vect{r}} d^3\vect{q}}  .
\end{equation}

A useful decomposition is
\begin{equation}
3\frac{ (\vect{\sigma_1}\cdot\vect{q}) (\vect{\sigma_2}\cdot\vect{q}) }{\mq^2 \pm \mu^2} = (\vect{\sigma_1}\cdot\vect{\sigma_2}) \mp (\vect{\sigma_1}\cdot\vect{\sigma_2}) \frac{\mu^2}{\mq^2 \pm \mu^2} + S_{12}(\vect{\hat{q}}) \frac{\mq^2}{\mq^2 \pm \mu^2}
\end{equation}
with $S_{12}(\vect{\hat{q}}) \equiv 3(\vect{\sigma}_1 \cdot \vect{\hat{q}})(\vect{\sigma}_2 \cdot \vect{\hat{q}}) - (\vect{\sigma}_1 \cdot \vect{\sigma}_2)$.

The form factor is $|F(\mq^2)|^2 \equiv \left(\frac{\Lambda^2 - m_{\pi}^2}{\mq^2 + X^2}\right)^2$ with $X^2 \equiv \Lambda^2 + \mu^2 - m_{\pi}^2 = \Lambda^2 - \tmu^2 - m_{\pi}^2$ and $\tmu^2 \equiv - \mu^2$.

\begin{table}[htb]
\begin{center}
\begin{tabular}{|c|c|}
\hline
$\bm{V(\vect{q})}$ & $\bm{V(\vect{r})}$  \\
\hline

$1$ & $\delta(\vect{r})$  \\
\hline

$\frac{1}{\mq^2 + \mu^2}$ & $\frac{1}{2\pi^2}\int_{0}^{\infty}{\frac{\mq^2 j_0(\mq r)}{\mq^2 + \mu^2}d\mq}$  \\
 & $\frac{1}{4\pi} \frac{e^{-\mu r}}{r}$  \\
 & $\frac{1}{4\pi} \frac{e^{ i \tmu r}}{r}$  \\
\hline

$S_{12}(\vect{\hat{q}})\frac{\mq^2}{\mq^2 + \mu^2}$ & $-S_{12}(\vect{\hat{r}}) \frac{1}{2\pi^2} \int_{0}^{\infty}{\frac{\mq^4 j_2(\mq r)}{\mq^2 + \mu^2}d\mq}$  \\
 & $-S_{12}(\vect{\hat{r}}) \frac{\mu^2}{4\pi} \frac{e^{- \mu r}}{r} \left(1 + \frac{3}{\mu r} + \frac{3}{(\mu r)^2} \right)$  \\
 & $S_{12}(\vect{\hat{r}}) \frac{\tmu^2}{4\pi} \frac{e^{i \tmu r}}{r} \left(1 + \frac{3 i}{\tmu r} - \frac{3}{(\tmu r)^2} \right)$  \\
\hline

$|F(\mq^2)|^2$ & $(\Lambda^2 - m_{\pi}^2)^2 \frac{1}{8 \pi} \frac{e^{-Xr}}{X}$  \\
\hline

$\frac{|F(\mq^2)|^2}{\mq^2 + \mu^2}$ & $\frac{1}{2\pi^2}\int_{0}^{\infty}{\frac{\mq^2 |F(\mq^2)|^2 j_0(\mq r)}{\mq^2 + \mu^2}d\mq}$  \\
 & $\frac{1}{4\pi} \left[ \frac{e^{-\mu r}}{r} - \frac{e^{-X r}}{r} - \frac{(\Lambda^2 - m_{\pi}^2)}{2X}{e^{-Xr}} \right]$  \\
 & $\frac{1}{4\pi} \left[ \frac{e^{i \tmu r}}{r} - \frac{e^{-X r}}{r} - \frac{(\Lambda^2 - m_{\pi}^2)}{2X}{e^{-Xr}} \right]$  \\
\hline

$S_{12}(\vect{\hat{q}})\frac{\mq^2|F(\mq^2)|^2}{\mq^2 + \mu^2}$ & $-S_{12}(\vect{\hat{r}}) \frac{1}{2\pi^2} \int_{0}^{\infty}{\frac{\mq^4 |F(\mq^2)|^2 j_2(\mq r)}{\mq^2 + \mu^2}d\mq}$  \\
 & $-S_{12}(\vect{\hat{r}}) \frac{1}{4\pi} \left[ \mu^2 \frac{e^{- \mu r}}{r} \left(1 + \frac{3}{\mu r} + \frac{3}{(\mu r)^2} \right) - X^2 \frac{e^{-X r}}{r} \left(1 + \frac{3}{X r} + \frac{3}{(X r)^2} \right) - (\Lambda^2 - m_{\pi}^2)\frac{e^{-Xr}}{2r}(1 + Xr) \right]$  \\
 & $-S_{12}(\vect{\hat{r}}) \frac{1}{4\pi} \left[ -\tmu^2 \frac{e^{i \tmu r}}{r} \left(1 + \frac{3i}{\tmu r} - \frac{3}{(\tmu r)^2} \right) - X^2 \frac{e^{-X r}}{r} \left(1 + \frac{3}{X r} + \frac{3}{(X r)^2} \right) - (\Lambda^2 - m_{\pi}^2)\frac{e^{-Xr}}{2r}(1 + Xr) \right]$  \\

\hline
\end{tabular}
\end{center}
\caption{Summary of Fourier transforms used}
\label{Table:FourierTransforms}
\end{table}

\section{Deuteron Potential}
\label{sec:Deuteron}

The deuteron is a combination of two nucleons in an isosinglet state with $J=1$, $S=1$, $L=0$ or $2$.  Because we have to sum over interactions between all the light quarks, relative to the heavy-light mesons the spin factors are changed to $-\frac{25}{3} \left( \begin{array}{cc} 1 & 0 \\ 0 & 1 \end{array} \right)$ for the central term and $-\frac{25}{3}\left( \begin{array}{cc} 0 & \sqrt{8} \\ \sqrt{8} & -2 \end{array} \right)$ for the tensor term\cite{Tornqvist:1993ng}.  The mixing between the $L=0$ and $L=2$ states is important in binding the deuteron.

If we follow Tornqvist and assume that pion exchange is the only binding mechanism, we can require the binding energy to be $\approx 2.22\ \MeV$ and so fix the scale $\Lambda$.  We take $V_0 = 1.3\ \MeV$ throughout.

We use three different expressions for the potential:
\begin{enumerate}
\item Our full expression given in Equ.\ \ref{Equ:FormFact:spin:mup}
\item Our expression without the `$\delta$ function' piece, Equ.\ \ref{Equ:FormFact:spin:nodeltafn:mup}
\item Tornqvist's expression\cite{Tornqvist:1993ng}
\end{enumerate}

We solve the potential by discretising the time-independent Schrodinger equation and then diagonalising the resulting matrix.  We take $N=500$ or $N=1000$ points for each $L$ and use maximum radii $R_0 = 0.10\ \MeV^{-1}$ and $0.15\ \MeV^{-1}$ to check for finite volume effects.  We set $\mu = m_{\pi} = 135\ \MeV$ and the reduced mass $\mu = 1/(1/m_n + 1/m_p)$ with $m_n = 939.57\ \MeV$ and $m_p = 938.27\ \MeV$.

Using potential (1) we obtain a binding energy $E=2.23\ \MeV$ with $\Lambda = 962\ \MeV$ and we find a relative amplitude squared of $0.93$ in $L=0$ and $0.07$ in $L=2$ with a RMS radius in $L=0$ of $0.02\ \MeV^{-1}$.

Using potential (2) we obtain a binding energy $E=2.20\ \MeV$ with $\Lambda = 752\ \MeV$ and we find a relative amplitude squared of $0.94$ in $L=0$ and $0.06$ in $L=2$ with a RMS radius in $L=0$ of $0.02\ \MeV^{-1}$.

Using potential (3) we obtain a binding energy $E=2.22\ \MeV$ with $\Lambda = 760\ \MeV$ and we find a relative amplitude squared of $0.94$ in $L=0$ and $0.06$ in $L=2$ with a RMS radius in $L=0$ of $0.02\ \MeV^{-1}$.

Although the results are sensitive to $\Lambda$, once this is fixed the other properties do not depend strongly on the details of the potential.

\section{Tables of Results}
\label{sec:ResultsTables}

\begin{table}[tb]
\begin{center}
\begin{tabular}{|c|c|c|c|c|c|c|c|c|}
\hline
\textbf{Potential} & \textbf{$\bm{\mu^2}$ Set} & $\bm{\Lambda / \MeV}$ & $\bm{E / \MeV}$ & $\bm{00,L=0}$ & $\bm{+-,L=0}$ & $\bm{00,L=2}$ & $\bm{+-,L=2}$ & \textbf{RMS Radius$\bm{/ \MeV^{-1}}$} \\
\hline
(1) & (A) and (B) & $800$ & (not bound) & & & & & \\
\hline
(1) & (A) & $962$ & $3870$ & $0.76$ & $0.22$ & $0.01$ & $0.01$ & $0.02$  \\
(1) & (B) & $962$ & $3870$ & $0.79$ & $0.19$ & $0.01$ & $0.01$ & $0.02$  \\ 
\hline
(1) & (A) & $1000$ & $3868$ & $0.69$ & $0.28$ & $0.02$ & $0.02$ & $0.01$  \\
(1) & (B) & $1000$ & $3869$ & $0.70$ & $0.27$ & $0.02$ & $0.02$ & $0.01$  \\ 
\hline
(1) & (A) & $1400$ & $3806$ & $0.50$ & $0.46$ & $0.02$ & $0.02$ & $0.003$  \\
(1) & (B) & $1400$ & $3807$ & $0.50$ & $0.46$ & $0.02$ & $0.02$ & $0.003$  \\ 
\hline
\hline
(2) & (A) and (B) & $752$ & (not bound) & & & & &  \\
\hline
(2) & (A) and (B) & $1500$ & (not bound) & & & & &  \\
\hline
(2) & (A) & $1750$ & $3871$ & $0.86$ & $0.11$ & $0.02$ & $0.02$ & $0.03$  \\
(2) & (B) & $1750$ & $3871$ & $0.90$ & $0.08$ & $0.01$ & $0.02$ & $0.03$  \\ 
\hline
(2) & (A) & $1800$ & $3870$ & $0.78$ & $0.17$ & $0.02$ & $0.03$ & $0.02$  \\
(2) & (B) & $1800$ & $3870$ & $0.81$ & $0.14$ & $0.02$ & $0.02$ & $0.02$  \\ 
\hline
(2) & (A) & $2000$ & $3863$ & $0.59$ & $0.33$ & $0.04$ & $0.04$ & $0.007$  \\
(2) & (B) & $2000$ & $3863$ & $0.60$ & $0.32$ & $0.04$ & $0.04$ & $0.008$  \\ 
\hline
\hline
(3) & (A) and (B) & $760$ & (not bound) & & & & &  \\
\hline
(3) & (A) and (B) & $1500$ & (not bound) & & & & &  \\
\hline
(3) & (A) & $1750$ & $3871$ & $0.86$ & $0.11$ & $0.02$ & $0.02$ & $0.03$  \\
(3) & (B) & $1750$ & $3871$ & $0.90$ & $0.08$ & $0.01$ & $0.01$ & $0.03$  \\ 
\hline
(3) & (A) & $2000$ & $3863$ & $0.59$ & $0.33$ & $0.04$ & $0.04$ & $0.007$  \\
(3) & (B) & $2000$ & $3863$ & $0.60$ & $0.32$ & $0.04$ & $0.04$ & $0.008$  \\ 
\hline
\end{tabular}
\end{center}
\caption{$D \bar{D}^*$ Results.  The amplitudes squared of the different components are given along with the RMS radius of the $00,L=0$ component.  The $+-$ and $00$ $L=0$ components have the same sign (i.e. would be an isoscalar state in the isospin symmetry limit).}
\label{Table:DDvResults}
\end{table}

\begin{table}[tb]
\begin{center}
\begin{tabular}{|c|c|c|c|c|c|c|c|}
\hline
\textbf{Potential} & $\bm{\Lambda / \MeV}$ & $\bm{E / \MeV}$ & $\bm{00,L=0}$ & $\bm{+-,L=0}$ & $\bm{00,L=2}$ & $\bm{+-,L=2}$ & \textbf{RMS Radius$\bm{/ \MeV^{-1}}$} \\
\hline
(1) & $800$ & $10580$ & $0.47$ & $0.48$ & $0.03$ & $0.03$ & $0.004$  \\
(1) & $962$ & $10540$ & $0.47$ & $0.48$ & $0.03$ & $0.03$ & $0.002$  \\
(1) & $1000$ & $10530$ & $0.47$ & $0.48$ & $0.03$ & $0.03$ & $0.002$  \\
(1) & $1400$ & $10270$ & $0.47$ & $0.47$ & $0.03$ & $0.03$ & $0.001$  \\
\hline
(2) & $752$ & (not bound) & & & & &  \\
(2) & $1000$ & $10600$ & $0.40$ & $0.50$ & $0.06$ & $0.05$ & $0.010$  \\
(2) & $1400$ & $10580$ & $0.41$ & $0.42$ & $0.08$ & $0.08$ & $0.004$  \\
(2) & $1750$ & $10520$ & $0.41$ & $0.41$ & $0.09$ & $0.09$ & $0.003$  \\
\hline
(3) & $760$ & (not bound) & & & & &  \\
(3) & $1000$ & $10600$ & $0.40$ & $0.49$ & $0.06$ & $0.05$ & $0.010$  \\
(3) & $1400$ & $10580$ & $0.41$ & $0.42$ & $0.08$ & $0.08$ & $0.004$  \\
(3) & $1750$ & $10520$ & $0.41$ & $0.41$ & $0.09$ & $0.09$ & $0.003$  \\
\hline
\end{tabular}
\end{center}
\caption{$B \bar{B}^*$ Results.  The amplitudes squared of the different components are given along with the RMS radius of the $00,L=0$ component.  The $+-$ and $00$ $L=0$ components have the same sign (i.e. would be an isoscalar state in the isospin symmetry limit).}
\label{Table:BBvResults}
\end{table}

\begin{table}[tb]
\begin{center}
\begin{tabular}{|c|c|c|c|c|c|c|}
\hline
\textbf{Potential} & $\bm{\Lambda / \MeV}$ & $\bm{E / \MeV}$ & $\bm{0+,L=0}$ & $\bm{+0,L=0}$ & $\bm{0+,L=2}$ & $\bm{+0,L=2}$  \\
\hline
(1) & $1700$ & (not bound) & & & &  \\ 
(1) & $2000$ & $3869$ & $0.46$ & $0.36$ & $0.09$ & $0.09$  \\ 
\hline
(2) & $1000$ & (not bound) & & & &  \\ 
(2) & $1500$ & $3871$ & $0.49$ & $0.39$ & $0.06$ & $0.06$  \\ 
(2) & $2000$ & $3814$ & $0.39$ & $0.38$ & $0.12$ & $0.12$  \\ 
\hline
(3) & $1000$ & (not bound) & & & &  \\ 
(3) & $1500$ & $3871$ & $0.49$ & $0.39$ & $0.06$ & $0.06$  \\ 
(3) & $2000$ & $3814$ & $0.39$ & $0.38$ & $0.12$ & $0.12$  \\ 
\hline
\end{tabular}
\end{center}
\caption{Exotic $D D^*$ Results.  The amplitudes squared of the different components are given.  The $0+$ and $+0$ $L=0$ components have the opposite sign (i.e. would be an isoscalar state in the isospin symmetry limit).}
\label{Table:ExoticDDvResults}
\end{table}

\begin{table}[tb]
\begin{center}
\begin{tabular}{|c|c|c|c|c|c|c|}
\hline
\textbf{Potential} & $\bm{\Lambda / \MeV}$ & $\bm{E / \MeV}$ & $\bm{0+,L=0}$ & $\bm{+0,L=0}$ & $\bm{0+,L=2}$ & $\bm{+0,L=2}$  \\
\hline
(1) & $700$ & (not bound) & & & &  \\ 
(1) & $1000$ & $10600$ & $0.33$ & $0.38$ & $0.15$ & $0.15$  \\ 
(1) & $1500$ & $10560$ & $0.27$ & $0.28$ & $0.23$ & $0.23$  \\ 
\hline
(2) & $500$ & (not bound) & & & &  \\ 
(2) & $700$ & $10600$ & $0.38$ & $0.46$ & $0.08$ & $0.08$  \\ 
(2) & $1000$ & $10590$ & $0.35$ & $0.36$ & $0.15$ & $0.15$  \\ 
\hline
(3) & $500$ & (not bound) & & & &  \\ 
(3) & $700$ & $10600$ & $0.38$ & $0.47$ & $0.08$ & $0.08$  \\ 
(3) & $1000$ & $10590$ & $0.35$ & $0.36$ & $0.15$ & $0.15$  \\ 
\hline
\end{tabular}
\end{center}
\caption{Exotic $B B^*$ Results.  The amplitudes squared of the different components are given.  The $0+$ and $+0$ $L=0$ components have the opposite sign (i.e. would be an isoscalar state in the isospin symmetry limit).}
\label{Table:ExoticBBvResults}
\end{table}

\end{document}